\documentstyle[12pt]{article}
\def\beq{\begin{equation}}
\def\eeq{\end{equation}}
\def\bea{\begin{eqnarray}}
\def\eea{\end{eqnarray}}
\def\bef{\begin{figure}}
\def\enf{\end{figure}}

\def\C{{\bf C}}
\def\Z{{\bf Z}}
\def\R{{\bf R}}
\def\N{{\bf N}}
\def\M{{\bf M}}
\def\CC{{\cal C}}
\def\CM{{\cal M}}
\def\CG{{\cal G}}

\def\ba{\begin{array}}
\def\ea{\end{array}}
\def\bce{\begin{center}}
\def\ece{\end{center}}

\def\Ga{\Gamma}

\def\vol#1{{\bf #1}}
\def\plb#1#2#3{Phys. Lett. {\bf B}\vol{#1#2#3} }
\def\nuphb#1#2#3{Nucl. Phys. {\bf B}\vol{#1#2#3} }

\newcommand{\drawsquare}[2]{\hbox{%
\rule{#2pt}{#1pt}\hskip-#2pt
\rule{#1pt}{#2pt}\hskip-#1pt
\rule[#1pt]{#1pt}{#2pt}}\rule[#1pt]{#2pt}{#2pt}\hskip-#2pt
\rule{#2pt}{#1pt}}

\newcommand{\fund}{\raisebox{-.5pt}{\drawsquare{6.5}{0.4}}}
\newcommand{\antifund}{\overline{\fund}}

\begin{document}
\begin{titlepage}
\rightline{BROWN-HET-1183}
\rightline{hep-th/9906012}
\vskip 1cm
\centerline{ \Large\bf{Branes at Orbifolded Conifold Singularities}}
\vskip 1cm
\centerline{\Large\bf{and Supersymmetric Gauge Field Theories}}
\vskip 1cm
\centerline{\sc Kyungho Oh$^{a}$ and Radu
Tatar$^{b}$}
\vskip 1cm
\centerline{ $^a$ Dept. of Mathematics, University of 
Missouri-St. Louis,
St. Louis, MO 63121, USA }
\centerline{{\tt oh@math.umsl.edu}}
\centerline{$^b$ Dept. of Physics, Brown University,
Providence, RI 02912, USA}
\centerline{\tt tatar@het.brown.edu}
\vskip 2cm
\centerline{\sc Abstract}
We consider D3 branes at orbifolded conifold singularities which are not
quotient singularities.
We use
toric geometry and gauged linear sigma model 
to study the moduli space of the gauge theories on the
D3 branes. We find that topologically distinct phases are related by
a flop transition. It is also shown 
that an orbifold
singularity can occur in some phases if we  give
expectation values to some of the chiral fields.
\def\today{\ifcase\month\or
January\or February\or March\or April\or May\or June\or
July\or August\or September\or October\or November\or December\fi,
\number\year}
\vskip 1cm
\end{titlepage}
\newpage
\section{Introduction}
\setcounter{equation}{0}
Last years have witnessed great insights into understanding of supersymmetric
 gauge
theory  and supergravity theory. We now found that these  are 
complementary descriptions of a single theory on  solitonic brane
solutions of M theory and string theory.
Configurations containing NS fivebranes 
and D branes in string theory are tools for studying supersymmetric 
gauge field theory in various dimensions with different supersymmetries (see \cite{gku} for a  
complete set of references up to February 1998). 

On the other hand, Maldacena's conjecture proposes that M or string theory
on the $AdS_{p} \times S^k$, with N units of supergravity $k$-form field 
through 
$S^{k}$ is dual to a $p - 1$ specific conformal field on the boundary of
the $AdS_{p}$ space \cite{mal} (see \cite{mal1} for an extensive review
and a complete set of references). 
The initial proposal gave conformal field theories with maximal
supersymmetry, ${\cal  N} = 4$ in four dimensions. This was obtained by
studying D3-branes in flat space. An immediate generalization  to
D3-branes at orbifold singularities breaks more supersymmetry
\cite{kac,law}. 

Another important class is obtained by studying D3-branes at non-orbifold
singularities like conifold singularity. The conifold singularity has been
analyzed in \cite{kw} 
where an infrared theory on the worldvolume of D3 branes was proposed. 
Other results for the case of non-orbifold singularities and their
connection to field theories in three and four dimensions have been obtained in 
\cite{ura,dm,afhs,gr1,mr,ot,bg,lopez,unge,karch,kw1,ahn,gk,agata,dia,gns}. 

In \cite{ura,dm}, the authors have exploited the fact that 
the conifold singularity is dual to a 
system of perpendicular NS5 fivebranes intersecting over a 3+1 dimensional
world-volume. Their result was generalized in \cite{unge,karch} for more
general conifolds. A duality between D3 branes on these general conifolds
and configurations of NS and D4 branes was proposed together with relation
between different resolutions of the singularity and displacements of NS
branes. In \cite{karch}, a mirror symmetry was proposed between
orbifolded conifolds and generalized conifolds.  

In the present paper we consider branes at orbifolded conifolds
$\CC_{k l}$ which is an orbifold of the three dimensional
conifold $xy-uv=0$ by a discrete group $\Z_k \times \Z_l$.
We show that the Higgs
branch of the moduli space of the gauge theory is the resolved or
(partially) resolved conifold singularity, depending on the values of
the FI parameters as holomorphic quotients.
The moduli spaces for  ${\cal N} = 2$ theories has been interpreted in terms of
symplectic quotients in a linear sigma model approach in \cite{witten},
and in terms of holomorphic quotients  in the mathematical approach
in \cite{agm}.
In \cite{dgm}, the latter approach has been extended to  ${\cal N} = 1$ theories
utilizing some ideas from \cite{agm, dmo}.
We also use toric geometry to study in detail the correspondence between D3
branes at orbifolded conifolds and brane configurations obtained after
T-dualities (For details on toric geometry see \cite{agm}).
In \cite{muto,gre,mra},  
D-branes on various other singularities have been studied in the lines of 
\cite{dgm}. The paper of \cite{unge} dealt with generalized conifolds 
$\CG_{k l}: x y = u^k v^l$.

The content of the paper is as follows: in section 2 we give a toric
description of the quotient singularity of the three dimensional conifold.
In section 3 we review  relevant field theory and  brane configurations.
We argue why brane box model is more suitable for orbifolded conifolds. 
In section 4 we derive  toric data for the simplest orbifolded conifold
$\CC_{2,2}$. In section 5 we derive toric data for the orbifolded
conifold $\CC_{2, 3}$. In section 6  we describe different phases of the
vacuum moduli space.

\section{Toric Geometry of Orbifolded Conifolds}
In this section, we will briefly  review toric singularity and its
physical realization by
the moduli space of the D-brane world-volume gauge theory  
on it via gauged linear sigma models to fix notations and terminologies. 
For detailed review, we refer to \cite{dgm}.
A toric variety is a space which contains algebraic torus ${(\C^*)}^d$ as an
 open dense subset. For example, a projective space 
${\bf P}^d =  (\C^{d+1} -\{0\})/\C^*$ is a toric variety because it contains
$(\C^*)^d \cong (\C^*)^{d+1}/\C^* \subset (\C^{d+1} -\{0\})/\C^*$.
As in the case of the projective space, we will express our toric varieties as
a quotient space (this can be thought of as a holomorphic quotient in 
the sense of the Geometric Invariant Theory~\cite{mfk}
 or as
a symplectic reduction as in gauged linear sigma model. In our cases, these two
will be the same \cite{kir}.):
\bea
V_\Delta = (\C^q - F_\Delta)//(\C^*)^{q-d}
\eea
where $q, F_\Delta$ and the action of $(\C^*)^{q-d}$ on $C^q$ are determined
by a combinatorial data $\Delta$. 
Now we give a description of the combinatorial data $\Delta$
for  Gorenstein canonical
singularity (i.e. a singularity with a trivial canonical class, $K$).
Consider  vectors  $v_1, \ldots , v_q$ in a lattice $\N = \Z^d \subset 
\N_{\R}=\N \otimes \R =\R^d$ in general position. 
We introduce the corresponding homogeneous coordinates $x_i$ for
of  $\C^q- F_\Delta$ in the holomorphic quotients. In gauged linear sigma
model, these correspond to  matter multiplets.
 There will be $(q-d)$ independent relations
\bea
\sum_{i=1}^q Q_i^{(a)}v_i = 0, \quad a=1, \ldots , q-d.
\eea
Here $Q^{(a)}$'s correspond to the charges of the matter fields under 
$U(1)^{q-d}$ which is the maximal compact subgroup of $(\C^*)^{q-d}$.  
The D-term equations will be 
\bea
\sum_{i=1}^{q} Q^{(a)}_i |x_i|^2 = r_a, \quad a =1, \ldots , q-d.
\eea
In the holomorphic quotient, the charge matrix whose column vectors
consist of  $Q^{(a)}$ 
determines the  action of $(\C^*)^{q-d}$ on $\C^q$ i.e. the action of
$(\lambda_1, \lambda_2, \ldots , \lambda_{q-d}) \in (\C^*)^{q-d}$ on
$(x_1, \ldots , x_q) \in \C^q$ is given by
\bea
(\lambda_1^{Q_1^{(1)}} 
\lambda_2^{Q_1^{(2)}}
\cdots  
\lambda_{q-d}^{Q_1^{(q-d)}}x_1, 
\lambda_1^{Q_2^{(1)}} 
\lambda_2^{Q_2^{(2)}}\cdots  
\lambda_{q-d}^{Q_2^{(q-d)}}x_2, 
\ldots ,
\lambda_1^{Q_q^{(1)}} 
\lambda_2^{Q_q^{(2)}}
\cdots  
\lambda_{q-d}^{Q_q^{(q-d)}}x_q)
\eea
Here the action can be carried out
as written or in two steps, an $(\R_+)^{q-d}$ action and a $U(1)^{q-d}$ action
if K\"ahler. The quotient will depend on the gauge fixing determined by
the $(\R_+)^{q-d}$ action i.e. the moment map. In the holomorphic approach,
this corresponds to different spaces $F_\Delta$ 
which give rise to (partial) resolutions of
the original space $V_\Delta$. 
In toric diagram, this corresponds to different
triangulations of a convex cone in $\R^d$ determined
by $\{ v_1, \ldots , v_q\}$. The collection of these combinatorial
data is  denoted by  $\Delta$ called a fan. 
The quotient space $V_\Delta$ 
will have Gorenstein canonical singularity if there exists
$u \in \Z^d$ such that $u \cdot v_i =1$ for all $i$ \cite{reid}. 
Thus $v_i$'s will
lie on the hyperplane with normal $u$ at a distance $1/\|u \|$ 
from the origin in $\R^d$.
This imposes the following condition on the charge vectors $Q^{(a)}$:
\bea
\sum_{i=1}^q Q^{(a)}_i = 0, \quad a=1, \ldots , q-d
\eea

To put our discussions in the language of the gauged linear sigma model,
recall that $\C^q$ is a symplectic manifold with the standard symplectic
form $\omega = i\sum_{i=1}^q dz^i \wedge d\bar{z}^{\bar{i}}$. The maximal
compact subgroup $G:=U(1)^{q-d}$ of $(\C^*)^{q-d}$ acts covariantly
on a symplectic manifold $(\C^q, \omega)$ by symplectomorphisms. 
The infinitesimal action will give rise to a moment map $\mu: \C^q \to 
{\bf g}^*$ by Poisson brackets.
In coordinates, the components of $\mu : \C^q \to \R^{q-d}$ are given by
\bea
\label{moment}
\mu_a = \sum_{i=1}^q Q^{(a)}_i |x_i|^2 - r_a
\eea
where $r_a$ are undetermined additive constants. The symplectic reduction
is then defined as
\bea
V(r) \equiv \mu^{-1}(0)/G.
\eea
The structure of $V(r)$ will depend on $r$. It follows from (\ref{moment})
that every $(\C^*)^{q-d}$-orbit in $\C^q$ will contribute at most one point
to $V(r)$. The value of $r$ will determine
the contributing orbits. For a fixed  $r$,
the set of $(\C^*)^{q-d}$-orbits which do not contribute is precisely
$F_\Delta$. The quotient space $V(r)$ carries a symplectic form $\omega_r$
by reducing $\omega$. The symplectic reduction carries a natural complex
structure, in which the reduced symplectic form becomes a K\"ahler form.

Now we will consider quotient singularities of the conifold (i.e. orbifolded
conifold). The conifold is a three dimensional hypersurface singularity in 
$\C^4$ defined by:
\bea
{\cal C}: \quad xy -uv = 0.
\eea
The conifold can be realized as a holomorphic quotient of $\C^4$
by the $\C^*$ action given by \cite{witten, kw}
\bea
(A_1, A_2,B_1, B_2)\mapsto (\lambda A_1, \lambda A_2,\lambda^{-1} B_1,
\lambda^{-1} B_2)\quad\mbox{ for }\lambda \in \C^*.
\eea
Thus the charge matrix is the transpose of $Q^{'}
=(1,1,-1,-1)$ and $\Delta=\sigma$ will be a convex polyhedral cone
in $\N^{'}_{\R}=\R^3$ 
generated by $v_1, v_2, v_3, v_4 \in \N^{'}=\Z^3$  where
\bea
v_1=(1,0,0), \quad v_2=(0,1,0),\quad  v_3=(0,0,1),\quad
v_4=(1,1,-1).
\eea
The isomorphism between the conifold ${\cal C}$ and the holomorphic
quotient is given by
\bea
\label{act}
x=A_1B_1, \quad y=A_2B_2, \quad u=A_1B_2, \quad v=A_2B_1. 
\eea
We take a further quotient of the conifold ${\cal C}$ by a discrete group
$\Z_k \times \Z_l$. Here $\Z_k$ acts on $A_i, B_j$ by
\bea
\label{zk}
(A_1, A_2, B_1, B_2) \mapsto 
(e^{-2\pi i/k} A_1, A_2, e^{2\pi i/k}B_1, B_2),
\eea
and $\Z_l$ acts by
\bea
\label{zl}
(A_1, A_2, B_1, B_2) \mapsto 
(e^{-2\pi i/l} A_1, A_2, B_1, e^{2\pi i/l}B_2).
\eea
Thus they will act on the conifold ${\cal C}$ by
\bea
\label{xy}
(x,y,u,v) \mapsto (x,y,e^{-2\pi i/k}u, e^{2\pi i/k}v)
\eea
and
\bea 
\label{uv}
(x,y,u,v) \mapsto (e^{-2\pi i/l}x,e^{2\pi i/l}y, u, v).
\eea
Its quotient is  called the hyper-quotient of the conifold
or the orbifolded conifold and denoted by ${\cal C}_{kl}$.
To put the actions (\ref{act}), (\ref{zk}) and (\ref{zl}) on an equal footing,
consider the over-lattice $\N$:
\bea
\N = \N^{'} + \frac{1}{k}(v_3-v_1) + \frac{1}{l}(v_4 -v_1).
\eea
Now the lattice  points $\sigma \cap \N$ of $\sigma$ in $\N$
 is generated by
$(k+1)(l+1)$ lattice points as a semigroup (These lattice points will be
referred as a toric diagram.). The charge matrix $Q$  will be
$(k+1)(l+1)$ by $(k+1)(l+1)-3$. The discrete group $\Z_k \times \Z_l \cong 
\N / \N^{'}$ will act on the conifold $\C^4 // U(1)$ and its quotient
will be the symplectic reduction $\C^{(k+1)(l+1)} // U(1)^{(k+1)(l+1)-3}$
with the moment map associated with  the charge matrix $Q$. The new toric 
diagram for ${\cal C}_{kl}$ will also lie
on the plane at a distance $1/\sqrt{3}$ from the origin with a normal vector
$(1,1,1)$ and we draw a toric diagram on the plane for ${\cal C}_{57}$:
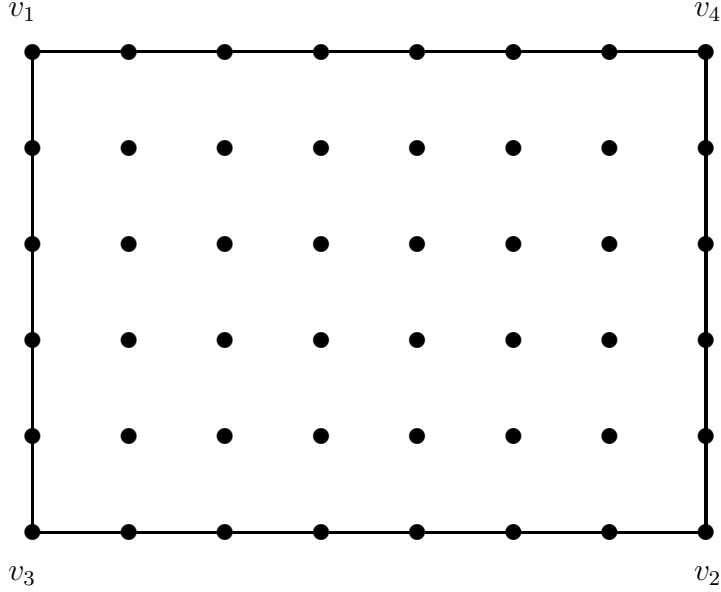
\begin{figure}
\setlength{\unitlength}{0.00083300in}%
\begingroup\makeatletter\ifx\SetFigFont\undefined%
\gdef\SetFigFont#1#2#3#4#5{%
  \reset@font\fontsize{#1}{#2pt}%
  \fontfamily{#3}\fontseries{#4}\fontshape{#5}%
  \selectfont}%
\fi\endgroup%
\begin{picture}(1000,3705)(2000,-4897)
\thicklines
\put(3601,-2161){\circle*{100}}
\put(3601,-2761){\circle*{100}}
\put(4201,-1561){\circle*{100}}
\put(3601,-3361){\circle*{100}}
\put(3601,-3961){\circle*{100}}
\put(4801,-1561){\circle*{100}}
\put(5401,-1561){\circle*{100}}
\put(6001,-1561){\circle*{100}}
\put(6601,-1561){\circle*{100}}
\put(7201,-1561){\circle*{100}}
\put(7201,-2161){\circle*{100}}
\put(7201,-2761){\circle*{100}}
\put(7201,-3361){\circle*{100}}
\put(7201,-3961){\circle*{100}}
\put(6601,-3961){\circle*{100}}
\put(6001,-3961){\circle*{100}}
\put(5401,-3961){\circle*{100}}
\put(4801,-3961){\circle*{100}}
\put(4201,-3961){\circle*{100}}
\put(4201,-3361){\circle*{100}}
\put(4801,-3361){\circle*{100}}
\put(5401,-3361){\circle*{100}}
\put(6001,-3361){\circle*{100}}
\put(6601,-3361){\circle*{100}}
\put(6601,-2761){\circle*{100}}
\put(6601,-2161){\circle*{100}}
\put(6001,-2161){\circle*{100}}
\put(6001,-2761){\circle*{100}}
\put(5401,-2161){\circle*{100}}
\put(5401,-2761){\circle*{100}}
\put(4801,-2161){\circle*{100}}
\put(4801,-2761){\circle*{100}}
\put(4201,-2161){\circle*{100}}
\put(4201,-2761){\circle*{100}}
\put(7801,-1561){\circle*{100}}
\put(7801,-2161){\circle*{100}}
\put(7801,-2761){\circle*{100}}
\put(7801,-3361){\circle*{100}}
\put(7801,-3961){\circle*{100}}
\put(3601,-1561){\circle*{100}}
\put(3601,-4561){\circle*{100}}
\put(4201,-4561){\circle*{100}}
\put(4801,-4561){\circle*{100}}
\put(5401,-4561){\circle*{100}}
\put(6001,-4561){\circle*{100}}
\put(6601,-4561){\circle*{100}}
\put(7201,-4561){\circle*{100}}
\put(7801,-4561){\circle*{100}}
\put(3451,-1336){$v_1$}
\put(7726,-1336){$v_4$}
\put(3451,-4861){$v_3$}
\put(7726,-4861){$v_2$}
\put(3601,-1561){\line( 0,-1){3000}}
\put(3601,-4561){\line( 1, 0){4200}}
\put(7801,-4561){\line( 0, 1){3000}}
\put(7801,-1561){\line(-1, 0){4200}}
\end{picture}

\caption{A toric diagram for ${\bf Z}_5\times {\bf Z}_7$ hyper-quotient 
of the conifold, ${\cal C}_{57}$}
\end{figure}
The action $(\ref{xy}), (\ref{uv})$
of $\Z_k \times \Z_l$ on the conifold ${\cal C}$ can be 
lifted to an action on $\C^4$ whose coordinates are $x,y, u, v$.
The ring of invariants will be $\C[x^l, y^l, xy, u^k, v^k, uv]$ and the
orbifolded conifold ${\cal C}_{kl}$ will be defined by the ideal
$(xy-uv)\C[x^l, y^l, xy, u^k, v^k, uv]$. Thus after renaming variables,
the defining equation for the orbifolded conifold will be
\bea
\label{con-eqn}
{\cal C}_{kl}: xy =z^l, \quad uv =z^k.
\eea
\section{Branes at the singularities  and Gauge Theory}
We now put branes to probe the geometric background space. Consider a system of
$M$ D3 branes sitting at the orbifolded conifold $\CC_{kl}$ in the
transversal direction. In the spirit of \cite{law}, 
the corresponding gauge field theory on the world-volume of
$D3$ has been obtained in \cite{ura} by investigating the action of 
the discrete group
on the field theory of the conifold developed in \cite{kw}.

The discrete group $\Z_k \times \Z_\l$ acts on the fields
$A_i, B_i$ of the conifold theory  as
in (\ref{zk}) and (\ref{zl}). 
By starting with a conifold theory with a
group $SU(klM) \times SU(klM)$,
 we obtain via the projection induced by the actions
the following ${\cal N}=1$ supersymmetric generically chiral
gauge theory for a specific choice for the Chan-Paton matrices:
\begin{equation}
\prod_{i=1}^{k}\prod_{j = 1}^{l} SU(M)_{i,j} \times
\prod_{i=1}^{k}\prod_{j = 1}^{l} SU(M)_{i,j}^{'}
\label{group}
\end{equation}
with matter content
\begin{center}
\begin{tabular}{ll}
{\bf Field} & {\bf Repr.} \\
$(A_1)_{i+1,j+1;i,j}$ & $(\fund_{i+1,j+1},\antifund^{'}_{i,j})$ \\
$(A_2)_{i,j;i,j}$ & $(\fund_{i,j},\antifund^{'}_{i,j})$ \\
$(B_1)_{i,j;i,j+1}$ & $(\fund^{'}_{i,j},\antifund_{i, j+1})$ \\
$(B_2)_{i,j;i+1,j}$ & $(\fund^{'}_{i,j},\antifund_{i+1,j})$
\end{tabular}
\end{center}
as explained in \cite{ura}. 
The superpotential is obtained by substituting the surviving fields into the 
conifold superpotential:
\bea
W = \sum_{i,j} (A_1)_{i+1,j+1;i,j} (B_1)_{i,j;i,j+1} (A_2)_{i,j+1;i,j+1}
(B_2)_{i,j+1;i+1,j+1}\\ \nonumber
\phantom{asf====}-
\sum_{i,j} (A_1)_{i+1,j+1;i,j} (B_1)_{i,j;i+1,j} (A_2)_{i+1,j;i+1,j}
(B_2)_{i+1,j;i+1,j+1}
\eea
Moreover, by giving a vev to all the fields $(A_2)_{i,j;i-1,j-1}$, we
obtain an $\prod_{i,j} SU(M)_{i,j}$ gauge theory with surviving chiral
multiplets $(A_1)_{i,j;i-1,j-1}, (B_1)_{i,j;i,j+1}, (B_2)_{i,j;i,j+1}$.
The superpotential for these fields will be
\bea
W= \sum_{i,j}(A_1)_{i,j;i-1,j-1}(B_1)_{i-1,j-1;i-1,j}
(B_2)_{i-1,j;i,j}\\ \nonumber
-(A_1)_{i,j;i-1,j-1}(B_1)_{i-1,j-1;i,j-1}
(B_2)_{i,j-1;i,j}
\eea
This field theory is that appearing on D3 branes on an orbifold 
$\C^3/\Z_k \times \Z_l$. 

We now discuss how to arrive from the configurations with D3 branes at 
conifold singularities to configurations with D4 or D5 branes together
with both types of NS branes.
>From (\ref{con-eqn}), we see that the orbifolded conifold can be viewed as a 
$\C^{*}\times \C^*$ fibration over the $z$ plane.
In other words, for
generic values of $z$, the pairs of variables $(x, y)$ and $(u, v)$ describe 
$\C^{*}\times\C^{*} $. 
Because we have  $\C^{*}\times\C^{*} $ fibration over the $z$ plane, 
we have two different kind $ U(1)$
orbits, one in each $\C^{*}$ fiber. So we can perform one T-duality or two
T-dualities along each of these orbits. 
If we make one T-duality we obtain a configuration with 
$k$ NS branes on a circle and all the configuration is at a $Z_{l}$
singularity. As first explained in \cite{lykken} and developed in
\cite{ura}, this is a chiral theory. Because we still have a singularity
which cannot be controlled by removing NS branes, it is more advantageous to
do both T-dualities in order to use all the geometrical information. 
By making these, we arrive to brane box configurations with two compact
direction, containing D5 branes together with both types of NS branes.
So by using the geometry, we study the K\"ahler deformation of the
orbifolded conifold with brane boxes. As explained in \cite{karch},
in order to account the number of K\"ahler structure parameters necessary to
completely solve the singularity, we need to modify 
the intersections of the NS branes by so-called 
 diamonds.
By closing the diamonds we turn off the B field and by rotating 
the diamonds on a plane perpendicular on the D5 brane we resolve the
singularity to $\C^3/\Z_k \times \Z_l$.

\section{The Orbifolded Conifold $\CC_{22}$ }
Consider a system of $M$ D3 branes sitting at the orbifolded conifold
\bea
{\CC}_{2 2} : x y =uv =z^2
\eea
As explained before, this is a chiral theory with
 the gauge group:
\begin{equation}
\prod_{i,j = 1}^{2} SU(M)_{i,j} \times \prod_{i,j = 1}^{2} SU(M)^{'}_{i,j}. 
\label{group1}
\end{equation}
Because the T-dual theory contains NS branes which are perpendicular, the
adjoint fields become massive and they are integrated out, leaving only
quadratic terms in the superpotential. For simplicity we denote the 16
fields by:
\bea 
\begin{array}{llll}
A_{11}= (A_1)_{22;11},& A_{12} = (A_1)_{21;12},&
A_{13} = (A_1)_{12;21},&  A_{14} = (A_1)_{11;22},\\ 
 A_{21} = (A_2)_{11;11},&
 A_{22} =  (A_2)_{12;12},&A_{23} = (A_2)_{21;21},&  
  A_{24} = (A_2)_{22;22},\\
B_{11} =  (B_1)_{11;12},&  B_{12} = (B_1)_{12;11},&
B_{13} =  (B_1)_{21;22},&
 B_{14} = (B_1)_{22;21},\\
  B_{21} = (B_2)_{11;21},&  B_{22} =
(B_2)_{12;22},&
  B_{23} = (B_2)_{21;11},&  B_{24} =  (B_2)_{22;12}.
\end{array}
\eea
The D term equations are:
\begin{eqnarray}
\label{dz2}
|A_{14}|^2 + |A_{21}|^2 - |B_{12}|^2 - |B_{23}|^2 = \xi_1  \\ \nonumber 
|A_{13}|^2 + |A_{22}|^2 - |B_{11}|^2 - |B_{24}|^2 = \xi_2  \\ \nonumber
|A_{12}|^2 + |A_{23}|^2 - |B_{14}|^2 - |B_{21}|^2 = \xi_3  \\ \nonumber
|A_{11}|^2 + |A_{24}|^2 - |B_{13}|^2 - |B_{22}|^2 = \xi_4  \\ \nonumber
|B_{21}|^2 + |B_{11}|^2 - |A_{11}|^2 - |A_{21}|^2 = \xi_5  \\ \nonumber
|B_{22}|^2 + |B_{12}|^2 - |A_{12}|^2 - |A_{22}|^2 = \xi_6  \\ \nonumber
|B_{23}|^2 + |B_{13}|^2 - |A_{13}|^2 - |A_{23}|^2 = \xi_7  \\ \nonumber
|B_{24}|^2 + |B_{14}|^2 - |A_{14}|^2 - |A_{24}|^2 = \xi_8  \
\end{eqnarray} 
 where the FI parameters satisfy the constraint
\begin{equation}
\sum_{i=1}^{8} \xi_i = 0
\end{equation}
The superpotential is
\begin{eqnarray}
\label{pot1}
W = A_{11} B_{11} A_{22} B_{22} + A_{12} B_{12} A_{21} B_{21} + A_{13} 
B_{13} A_{24} B_{24} + A_{14}
B_{14} A_{23} B_{23} - \\ \nonumber
- A_{11} B_{21} A_{23} B_{13} - A_{12} B_{22} A_{24} B_{14} - A_{13}
B_{23} A_{21} B_{11}
- A_{14} B_{24} A_{22} B_{12} \
\end{eqnarray}
There are 16 F-term constraints derived from this superpotential, not all
of them independent. As opposed to other field theories considered
previously in the literature, our case involves chiral fields so the 
F term equations will give equality between two products of three fields
as for example the one obtained after taking the derivative with
$A_{11}$ : $ B_{11} A_{22} B_{22} =  B_{21} A_{23} B_{13}$ and the rest of
15 equations are similar.  After solving the independent F-term
equations, 
we arrive at 10 independent
fields, the rest of 6 fields being expressed in terms of these. 
We chose $A_{24}, A_{13}, A_{14}, B_{12}, B_{13}, B_{14}, B_{21}, B_{22},
 B_{23}, B_{24}$ as the independent variables. The solution for the F-term
equations is:
\bea
\begin{array}{rrrrrrrrrrr}
&A_{24}& A_{13}& A_{14}& B_{12}& B_{13}& B_{14}& B_{21}& B_{22}& B_{23}& B_{24}\\
A_{11}&0&0&       1&      0&     -1&      1&     -1&     0&       1&    0\\
A_{12}&0&1&       0&      0&      1&     -1&      0&    -1&       0&    1\\
A_{21}&1&0&       0&     -1&      0&      1&     -1&     1&       0&    0\\
A_{22}&1&1&      -1&     -1&      1&      0&      0&     0&       0&    0\\
A_{23}&1&1&      -1&      0&      1&     -1&      0&     0&      -1&    1\\
B_{11}&0&0&       0&      1&      1&     -1&      1&    -1&      -1&    1
\end{array}
\eea
We now proceed to obtain the vacuum moduli space in the usual way,
i.e. by imposing the F-term constraints and the D-term constraints in the
form of symplectic quotients as the gauged linear sigma model. 
If we impose only F-term constraints,  we can identify the moduli, denoted by
$\CM_F$, of
the 16 fields as a cone $\M_+$ in $\M =\Z^{10}$ by expressing them in terms of
10 independent fields. To construct this as a symplectic quotient, we consider
the dual cone $\N_+$ in $\N =  \mbox{Hom } (\M, \Z)$. It turns out that
the dual cone $\N_+$ is generated by $24$ lattice points. 
Thus we have a map 
\bea
T: \Z^{24} \to \N,
\eea
which is shown in the Figure 2. 
\bce
\begin{figure} 
\label{T}
{\small
$$
\left(
\begin{array}{rrrrrrrrrrrrrrrrrrrrrrrr}
 0& 0& 1& 0& 0& 0& 0& 0& 0& 0& 1& 1& 0& 0& 0& 0& 0& 0& 1& 1& 1& 0& 0& 0\\ 
  0& 1& 0& 0& 0& 0& 0& 0& 0& 1& 0& 0& 0& 0& 0& 1& 1& 1& 0& 0& 0& 0& 0& 1\\ 
  0& 0& 0& 0& 0& 0& 0& 0& 1& 1& 0& 1& 0& 0& 1& 0& 0& 0& 0& 0& 1& 0& 0& 1\\ 
  0& 0& 0& 0& 0& 0& 0& 0& 0& 0& 1& 0& 0& 1& 0& 0& 1& 1& 0& 1& 0& 0& 1& 0\\ 
  0& 0& 0& 0& 0& 0& 1& 1& 1& 0& 0& 0& 0& 1& 1& 0& 0& 0& 0& 0& 0& 0& 1& 0\\ 
  0& 0& 0& 0& 0& 1& 0& 1& 0& 0& 0& 0& 1& 1& 0& 1& 0& 1& 0& 0& 0& 0& 0& 0\\ 
  0& 0& 0& 0& 0& 0& 0& 0& 0& 0& 0& 0& 1& 0& 0& 1& 0& 0& 1& 0& 1& 1& 0& 1\\ 
  0& 0& 0& 0& 1& 0& 0& 0& 0& 0& 0& 0& 0& 0& 1& 0& 1& 0& 0& 0& 0& 1& 1& 1\\ 
  0& 0& 0& 1& 0& 0& 1& 0& 0& 0& 0& 0& 0& 0& 0& 0& 0& 0& 1& 1& 0& 1& 1& 0\\ 
  1& 0& 0& 1& 1& 1& 0& 0& 0& 0& 0& 0& 1& 0& 0& 0& 0& 0& 0& 0& 0& 1& 0& 0
\end{array}
\right)
$$
}
\caption{The $10 \times24$ matrix $T$}
\end{figure}
\ece

The transpose of the kernel
of $T$ is then a $14 \times 24$ charge matrix $Q$ 
which is shown Figure 3. 
\bef\label{Q}
{\small
$$
\left(
\begin{array}{rrrrrrrrrrrrrrrrrrrrrrrr}
2& -1& 0& -1& -1& 1& 1& 0& -1& 0& 0& 0& -1& 0& 0& 0& 0& 0& 0& 0& 0& 0& 0& 
   1\\ 1& 0& 1& 0& -1& 0& -1& 0& 0& 0& -1& 0& 0& 0& 0& 0& 0& 0& 0& 0& 0& 0& 
   1& 0\\ 1& 0& 0& -1& -1& 1& 0& 0& 0& 0& 0& 0& -1& 0& 0& 0& 0& 0& 0& 0& 0& 
   1& 0& 0\\ 1& 0& -1& -1& 0& 1& 1& 0& -1& 0& 0& 0& -1& 0& 0& 0& 0& 0& 0& 
   0& 1& 0& 0& 0\\ 1& 0& 0& -1& 0& 0& 0& 0& 0& 0& -1& 0& 0& 0& 0& 0& 0& 0& 
   0& 1& 0& 0& 0& 0\\ 1& 0& -1& -1& 0& 1& 0& 0& 0& 0& 0& 0& -1& 0& 0& 0& 0& 
   0& 1& 0& 0& 0& 0& 0\\ 1& -1& 1& 0& 0& -1& 0& 0& 0& 0& -1& 0& 0& 0& 0& 0& 
   0& 1& 0& 0& 0& 0& 0& 0\\ 1& -1& 1& 0& -1& 0& 0& 0& 0& 0& -1& 0& 0& 0& 0& 
   0& 1& 0& 0& 0& 0& 0& 0& 0\\ 1& -1& 0& 0& 0& 0& 0& 0& 0& 0& 0& 0& -1& 0& 
   0& 1& 0& 0& 0& 0& 0& 0& 0& 0\\ 
  1& 0& 0& 0& -1& 0& 0& 0& -1& 0& 0& 0& 0& 0& 1& 0& 0& 0& 0& 0& 0& 0& 0& 0\\ 
  0& 0& 1& 1& 0& -1& -1& 0& 0& 0& -1& 0& 0& 1& 0& 0& 0& 0& 0& 0& 0& 0& 0& 
   0\\ 1& 0& -1& -1& 0& 0& 1& 0& -1& 0& 0& 1& 0& 0& 0& 0& 0& 0& 0& 0& 0& 0& 
   0& 0\\ 1& -1& 0& -1& 0& 0& 1& 0& -1& 1& 0& 0& 0& 0& 0& 0& 0& 0& 0& 0& 0& 
   0& 0& 0\\ 0& 0& 0& 1& 0& -1& -1& 1& 0& 0& 0& 0& 0& 0& 0& 0& 0& 0& 0& 0& 
   0& 0& 0& 0
\end{array}
\right)
$$}
\caption{The $14\times 24$ charge matrix $Q$}
\enf
Thus we have an exact sequence:
\bea
0 \longrightarrow \Z^{14} \stackrel{^{t}Q}
{\longrightarrow} \Z^{24} \stackrel{T}{\longrightarrow} \N \longrightarrow
 0.
\eea
>From this sequence, one can see that the moduli space $\CM_F$ can be expressed as
a holomorphic quotient of $\C^{24}$ by $(\C^*)^{14}$ whose action is 
specified by $Q$ (or a symplectic quotient by $U(1)^{14}$.) via the map 
induced by $T$.
To incooperate 
the D-term  constraints, we need to see how the action of $(\C^*)^{10}$ on  the
toric variety $\CM_F$ is represented in these terms. Since the action of 
$(\C^*)^{10}$ on the open subset $(\C^*)^{10}\subset \CM_F$ 
must be  the obvious multiplication, the action of $(\C^*)^{10}$ on $\C^{24}$
is specified the transpose of a $10\times 24$ matrix $U$ such that
\bea
T\cdot ^{t}\!U = \mbox{ Id}_k.
\eea
$U$ is shown in the Figure 4. 
\bef
\label{U}
{\small
$$
\left(
\begin{array}{rrrrrrrrrrrrrrrrrrrrrrrr}
0& 0& 1& 0& 0& 0& 0& 0& 0& 0& 0& 0& 0& 0& 0& 0& 0& 0& 0& 0& 0& 0& 0& 0\\ 
  0& 1& 0& 0& 0& 0& 0& 0& 0& 0& 0& 0& 0& 0& 0& 0& 0& 0& 0& 0& 0& 0& 0& 0\\ 
  -1& 0& 0& 1& 0& 0& -1& 0& 1& 0& 0& 0& 0& 0& 0& 0& 0& 0& 0& 0& 0& 0& 0& 0\\ 
  0& 0& -1& 0& 0& 0& 0& 0& 0& 0& 1& 0& 0& 0& 0& 0& 0& 0& 0& 0& 0& 0& 0& 0\\ 
  1& 0& 0& -1& 0& 0& 1& 0& 0& 0& 0& 0& 0& 0& 0& 0& 0& 0& 0& 0& 0& 0& 0& 0\\ 
  -1& 0& 0& 0& 0& 1& 0& 0& 0& 0& 0& 0& 0& 0& 0& 0& 0& 0& 0& 0& 0& 0& 0& 0\\ 
  0& 0& 0& 0& 0& -1& 0& 0& 0& 0& 0& 0& 1& 0& 0& 0& 0& 0& 0& 0& 0& 0& 0& 0\\ 
  -1& 0& 0& 0& 1& 0& 0& 0& 0& 0& 0& 0& 0& 0& 0& 0& 0& 0& 0& 0& 0& 0& 0& 0\\ 
  -1& 0& 0& 1& 0& 0& 0& 0& 0& 0& 0& 0& 0& 0& 0& 0& 0& 0& 0& 0& 0& 0& 0& 0\\ 
  1& 0& 0& 0& 0& 0& 0& 0& 0& 0& 0& 0& 0& 0& 0& 0& 0& 0& 0& 0& 0& 0& 0& 0
\end{array}
\right)
$$}
\caption{The $10\times 24$ matrix $U$}
\enf

The D-term equations are represented by
a matrix $V$ in
the Figure 5. 
\bef
\label{V}
{\small
$$
\left(
\begin{array}{rrrrrrrrrr}
0& 0& 1& -1& 0& 0& 0& 0& -1& 0\\ 0& 1& 0& 0& 0& 0& 0& 0& 0& -1\\ 
  0& 0& 0& 0& 0& -1& -1& 0& 0& 0\\ 1& 0& 0& 0& -1& 0& 0& -1& 0& 0\\ 
  0& 0& 0& 0& 0& 0& 1& 0& 0& 0\\ 0& 0& 0& 1& 0& 0& 0& 1& 0& 0\\ 
  0& -1& 0& 0& 1& 0& 0& 0& 1& 0
\end{array}
\right)
$$}
\caption{The $7\times 10$ matrix $V$}
\enf
\bef
{\small
$$
\left(
\begin{array}{rrrrrrrrrrrrrrrrrrrrrrrr}
2& -1& 0& -1& -1& 1& 1& 0& -1& 0& 0& 0& -1& 0& 0& 0& 0& 0& 0& 0& 0& 0& 0& 
   1\\ 1& 0& 1& 0& -1& 0& -1& 0& 0& 0& -1& 0& 0& 0& 0& 0& 0& 0& 0& 0& 0& 0& 
   1& 0\\ 1& 0& 0& -1& -1& 1& 0& 0& 0& 0& 0& 0& -1& 0& 0& 0& 0& 0& 0& 0& 0& 
   1& 0& 0\\ 1& 0& -1& -1& 0& 1& 1& 0& -1& 0& 0& 0& -1& 0& 0& 0& 0& 0& 0& 
   0& 1& 0& 0& 0\\ 1& 0& 0& -1& 0& 0& 0& 0& 0& 0& -1& 0& 0& 0& 0& 0& 0& 0& 
   0& 1& 0& 0& 0& 0\\ 1& 0& -1& -1& 0& 1& 0& 0& 0& 0& 0& 0& -1& 0& 0& 0& 0& 
   0& 1& 0& 0& 0& 0& 0\\ 1& -1& 1& 0& 0& -1& 0& 0& 0& 0& -1& 0& 0& 0& 0& 0& 
   0& 1& 0& 0& 0& 0& 0& 0\\ 1& -1& 1& 0& -1& 0& 0& 0& 0& 0& -1& 0& 0& 0& 0& 
   0& 1& 0& 0& 0& 0& 0& 0& 0\\ 1& -1& 0& 0& 0& 0& 0& 0& 0& 0& 0& 0& -1& 0& 
   0& 1& 0& 0& 0& 0& 0& 0& 0& 0\\ 
  1& 0& 0& 0& -1& 0& 0& 0& -1& 0& 0& 0& 0& 0& 1& 0& 0& 0& 0& 0& 0& 0& 0& 0\\ 
  0& 0& 1& 1& 0& -1& -1& 0& 0& 0& -1& 0& 0& 1& 0& 0& 0& 0& 0& 0& 0& 0& 0& 
   0\\ 1& 0& -1& -1& 0& 0& 1& 0& -1& 0& 0& 1& 0& 0& 0& 0& 0& 0& 0& 0& 0& 0& 
   0& 0\\ 1& -1& 0& -1& 0& 0& 1& 0& -1& 1& 0& 0& 0& 0& 0& 0& 0& 0& 0& 0& 0& 
   0& 0& 0\\ 0& 0& 0& 1& 0& -1& -1& 1& 0& 0& 0& 0& 0& 0& 0& 0& 0& 0& 0& 0& 
   0& 0& 0& 0\\ 0& 0& 1& 0& 0& 0& -1& 0& 1& 0& -1& 0& 0& 0& 0& 0& 0& 0& 0& 
   0& 0& 0& 0& 0\\ -1& 1& 0& 0& 0& 0& 0& 0& 0& 0& 0& 0& 0& 0& 0& 0& 0& 0& 
   0& 0& 0& 0& 0& 0\\ 1& 0& 0& 0& 0& 0& 0& 0& 0& 0& 0& 0& -1& 0& 0& 0& 0& 
   0& 0& 0& 0& 0& 0& 0\\ 0& 0& 1& 1& -1& 0& -1& 0& 0& 0& 0& 0& 0& 0& 0& 0& 
   0& 0& 0& 0& 0& 0& 0& 0\\ 0& 0& 0& 0& 0& -1& 0& 0& 0& 0& 0& 0& 1& 0& 0& 
   0& 0& 0& 0& 0& 0& 0& 0& 0\\ -1& 0& -1& 0& 1& 0& 0& 0& 0& 0& 1& 0& 0& 0& 
   0& 0& 0& 0& 0& 0& 0& 0& 0& 0\\ 
  0& -1& 0& 0& 0& 0& 1& 0& 0& 0& 0& 0& 0& 0& 0& 0& 0& 0& 0& 0& 0& 0& 0& 0
\end{array}
\right)
$$}
\caption{The $21\times 24$ matrix $\tilde{Q}$}
\enf
We ignored the charges on the dependent fields because
 they
are already encoded in $Q$. Thus on $\C^{24}$, the D-term constraints are 
represented by the charge matrix $VU$. Finally the full set of
charges is given by a $21\times 24$ charge matrix $\tilde{Q}$ (Figure 6.)
by  concatenating $Q$ and $VU$.
The cokernel of its  transpose gives toric data for our vacuum moduli space,
denoted by $\CM$. 
After eliminating redundant variables, it is give
in the form of a map $T_{\CM}: \Z^{9} \to \Z^3$:
\bea
T_{\CM}=\left(
\begin{array}{rrrrrrrrr}
2& 1& 0& 1& 0& -1& 0& -1& -2\\ 0& 1& 2& 0& 1& 2& 0& 1& 2\\ 
  -1& -1& -1& 0& 0& 0& 1& 1& 1
\end{array}
\right)
\eea
The lattice points given by $T_{\CM}$
lie on the plane with normal $(1,1,1)$
at a distance $1/\sqrt{3}$ from the origin. We depict these lattice points
\bea
\ba{lll}
v_{1} = (2, 0, -1),& v_{2} = (1, 1, -1),& v_{3} = (0, 2, -1), \\ 
v_{4} = (1, 0, 0),&  v_{5} = (0, 1, 0),& v_{6} = (-1, 2, 0), \\
v_{7} = (0, 0, 1),&  v_{8} = (-1, 1, 1),& v_{9} = (-2, 2, 1) 
\ea
\eea
in the planar diagram (Figure 7). This is exactly a toric diagram for
the orbifolded conifold $\CC_{22}: xy=uv=z^2$. 
The corresponding charge matrix $Q_{\CM}$
for the toric data $T_{\CM}$ with the Fayet-Iliopoulos D-term parameters
from (\ref{dz2}) is as follows:
\bea
\label{QM}
Q_{\CM}=\left(
\ba{cccccccccc}
0& 0& 0& 2& -2& 0& -1& 0& 1& 2\xi_1 + \xi_2 + \xi_4 + \xi_5 - \xi_7\\ 
  0& 0& 0& 1& -1& 0& -1& 1& 0& \xi_1 - \xi_7\\ 
  0& 0& 0& 1& -2& 1& 0& 0& 0& \xi_1 - \xi_3 - \xi_6 - \xi_7\\ 
   1& 0& 0& -2& 0& 0& 1& 0& 0& -\xi_1 - \xi_2 - \xi_5 - \xi_6\\ 
  0& 0& 1& 0& -2& 0& 1& 0& 0& -\xi_2 - \xi_3 - \xi_6 - \xi_7\\
0& 1& 0& -1& -1& 0& 1& 0& 0& -\xi_2 - \xi_6 
\ea
\right)
\eea
For this choice of redundant variables, Fayet-Iliopoulos D-term parameters
must satisfy
\bea
\label{fi-ieq}
\xi_1 > 0,\,\,\,
\xi_4 > 0,\,\,\, -\xi_6    > 0,\,\,\,
 - \xi_7 > 0,\,\,\,        
       -\xi_3      -\xi_6-\xi_7 > 0,\,\,\, \\   \nonumber
    -\xi_2         -\xi_6-\xi_7 > 0,\,\,\,    
   -\xi_2-\xi_3      -\xi_6-\xi_7 > 0,\,\,\,    
 -\xi_1-\xi_2      -\xi_5-\xi_6    > 0,\,\,\, \\   \nonumber
 -\xi_1-\xi_2-\xi_3   -\xi_5-\xi_6-\xi_7 > 0,\,\,\,    
   \xi_1      +\xi_4+\xi_5       > 0,\,\,\,    
 \xi_1+\xi_2   +\xi_4+\xi_5       > 0.  
\eea
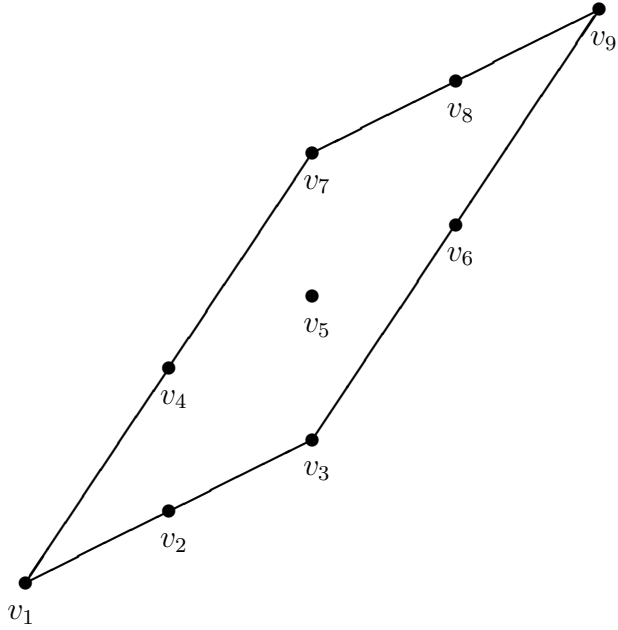
\begin{figure}
\setlength{\unitlength}{0.00063300in}%
\begingroup\makeatletter\ifx\SetFigFont\undefined%
\gdef\SetFigFont#1#2#3#4#5{%
  \reset@font\fontsize{#1}{#2pt}%
  \fontfamily{#3}\fontseries{#4}\fontshape{#5}%
  \selectfont}%
\fi\endgroup%
\begin{picture}(3,5210)(451,-5488)
\thicklines
\put(3601,-5161){\circle*{100}}
\put(4801,-4561){\circle*{100}}
\put(6001,-3961){\circle*{100}}
\put(7201,-2161){\circle*{100}}
\put(6001,-2761){\circle*{100}}
\put(4801,-3361){\circle*{100}}
\put(6001,-1561){\circle*{100}}
\put(7201,-961){\circle*{100}}
\put(8401,-361){\circle*{100}}
\put(6001,-1561){\line(-2,-3){2400}}
\put(3601,-5161){\line( 2, 1){2400}}
\put(6001,-3961){\line( 2, 3){2400}}
\put(8401,-361){\line(-2,-1){2400}}

\put(4726,-4861){$v_2$}
\put(3451,-5461){$ v_1$}
\put(5926,-4261){$v_3$}
\put(4726,-3661){$v_4$}
\put(5926,-3061){$v_5$}
\put(7126,-2461){$v_6$}
\put(5926,-1861){$v_7$}
\put(7126,-1261){$v_8$}
\put(8326,-661){$v_9$}
\end{picture}
\caption{${\bf Z}_2\times {\bf Z}_2$ orbifold of the conifold $\CC_{22}$}
\end{figure}
\section{The Orbifolded Conifold $x y = z^{2}, u v = z^3$}

In this case we start with a system of D branes sitting at the orbifold
conifold singularity
\bea
\CC_{23}:  
x y = z^{2}, u v = z^3.
\eea
By putting  on the $M$ D3 branes on $\CC_{23}$, we obtain  the field theory
with the gauge group:
\begin{equation}
\label{groupz3}
\prod_{i=1}^{2}\prod_{j=1}^{3} SU(M)_{i,j} \times 
\prod_{i=1}^{2}\prod_{j=1}^{3}
SU(M)^{'}_{i,j}
\end{equation}
The matter content for the theory with the gauge group (\ref{groupz3}) is
similar to the one encountered for the previous orbifolded conifold but we
have 24 fields now instead of 16 as before.
For simplicity we denote the 24 
fields by: 
\bea
\begin{array}{llll}
A_{11}= (A_1)_{22;11},&  A_{12} = (A_1)_{21;12},&
A_{13} = (A_1)_{32;21},& A_{14} = (A_1)_{31;22}\\ 
A_{15} = (A_1)_{12;31},& A_{16} = (A_1)_{11;32},&
A_{21} = (A_2)_{11;11},&
 A_{22} =  (A_2)_{12;12}\\ 
A_{23} = (A_2)_{21;21},&  
  A_{24} = (A_2)_{22;22},& A_{25} = (A_2)_{31;31},& A_{26} = (A_2)_{32;32}\\ 
B_{11} =  (B_1)_{11;12},&  B_{12} = (B_1)_{12;11},&
B_{13} =  (B_1)_{21;22},&
 B_{14} = (B_1)_{22;21}\\ 
 B_{15} = (B_1)_{31;32},& 
B_{16} = (B_1)_{32:31},& B_{21} = (B_2)_{11;21},&
B_{22} = (B_2)_{12;22}\\  
  B_{23} = (B_2)_{21;31},&  B_{24} =  (B_2)_{22;32},&
B_{25} = (B_2)_{31;11},& B_{26} = (B_2)_{32;12}
\end{array}
\eea
The superpotential is then:
\begin{eqnarray}
\label{pot2}
W = A_{11} B_{11} A_{22} B_{22} + A_{12} B_{12} A_{21} B_{21} + A_{13} 
B_{13} A_{24} B_{24} + A_{14}
B_{14} A_{23} B_{23} + \\ \nonumber
+ A_{15} B_{15} A_{26} B_{26} 
+ A_{16} B_{16} A_{25} B_{25} 
- A_{11} B_{21} A_{23} B_{13} - A_{12} B_{22} A_{24} B_{14} \\ \nonumber
- A_{13}
B_{23} A_{25} B_{15}
- A_{14} B_{24} A_{26} B_{16} - 
- A_{15} B_{25} A_{21} B_{11} - A_{16} B_{26} A_{22} B_{12} \
\end{eqnarray}
There are 24 F-term constraints derived from this superpotential, not all
of them independent and by solving them we arrive at 14 independent
fields, the rest of 10 fields being expressed in terms of these. 
We choose 
\bea
A_{16},\,\, A_{26},\,\, B_{11},\,\, B_{12},\,\, B_{13},\,\, B_{14},\,\, B_{15},\,\,
B_{16},\,\, B_{21},\,\, B_{22},\,\, B_{23},\,\, B_{24},\,\, B_{25},\,\, B_{26}
\eea
as the independent
fields.

The D term equations are:
\begin{eqnarray}
\label{dz3}
|A_{16}|^2 + |A_{21}|^2 - |B_{12}|^2 - |B_{25}|^2 = \xi_1  \\ \nonumber 
|A_{15}|^2 + |A_{22}|^2 - |B_{11}|^2 - |B_{26}|^2 = \xi_2  \\ \nonumber
|A_{12}|^2 + |A_{23}|^2 - |B_{14}|^2 - |B_{21}|^2 = \xi_3  \\ \nonumber
|A_{11}|^2 + |A_{24}|^2 - |B_{13}|^2 - |B_{22}|^2 = \xi_4  \\ \nonumber
|A_{14}|^2 + |A_{26}|^2 - |B_{16}|^2 - |B_{23}|^2 = \xi_5  \\ \nonumber
|A_{13}|^2 + |A_{25}|^2 - |B_{15}|^2 - |B_{24}|^2 = \xi_6  \\ \nonumber
|B_{21}|^2 + |B_{11}|^2 - |A_{11}|^2 - |A_{21}|^2 = \xi_7  \\ \nonumber
|B_{22}|^2 + |B_{12}|^2 - |A_{12}|^2 - |A_{22}|^2 = \xi_8  \\ \nonumber
|B_{23}|^2 + |B_{13}|^2 - |A_{13}|^2 - |A_{23}|^2 = \xi_9  \\ \nonumber
|B_{24}|^2 + |B_{14}|^2 - |A_{14}|^2 - |A_{24}|^2 = \xi_{10} \\ \nonumber
|B_{25}|^2 + |B_{15}|^2 - |A_{15}|^2 - |A_{25}|^2 = \xi_{11} \\ \nonumber
|B_{26}|^2 + |B_{16}|^2 - |A_{16}|^2 - |A_{26}|^2 = \xi_{12} 
\end{eqnarray}
where the FI parameters satisfy the constraint
\bea
\sum_{i=1}^{12} \xi_i = 0.
\eea

We want to implement the same procedure as in the previous $\Z_2 \times \Z_3$
orbifolded conifold $\CC_{22}$. As before, we can identify  the moduli space
$\CM_F$
of 24 fields under the F-term constraints as a cone $\M_+$ in $\M=\Z^{14}$.
The dual cone $\N_+$ is generated by 80 lattice points represented by $T$.  
Thus 
$\CM_F$ can be expressed as a symplectic quotient $\C^{80}// U(1)^{66}$ 
whose action is specified by $Q$. This can be expressed as the following 
exact sequence:
\bea
0 \longrightarrow \Z^{66} \stackrel{^{t}Q}
{\longrightarrow} \Z^{80} \stackrel{T}{\longrightarrow} \N \longrightarrow
 0.
\eea
By further imposing $11$ D-term equations from (\ref{dz3}), 
we obtain toric data for the vacuum moduli space $\CM$ as a three dimensional
toric variety $\C^{80}// U(1)^{77}$. Because of huge sizes of the
matrices involved, we only write the final toric data after 
eliminating redundant variables.
It is given in the form of a map $T_{\CM}: \Z^{12} \to \Z^{3}$:
\bea
T_{\CM}=\left(
\begin{array}{rrrrrrrrrrrr}
0&  1& 2& 3& -1& 0& 1&  2& -1&0&1&-2\\ 
0& -1&-2&-3&  2& 1& 0& -1&  3&2&1&4\\ 
1&  1& 1& 1&  0& 0& 0&  0& -1&-1&-1&-1
\end{array}
\right)
\eea
The lattice points given by $T_{\CM}$ are as follows:
\bea
\ba{llll}
v_1= (0, 0, 1),& v_2 =(1, -1, 1),& v_3  =(2, -2, 1),& v_4 =(3, -3, 1),\\ 
 v_5=(-1, 2, 0),& v_6=(0, 1, 0),& v_7 =(1, 0, 0),& v_8=(2, -1, 0),\\ 
v_9=(-1, 3, -1),& v_{10}=(0, 2, -1),&  v_{11}=(1, 1, -1),&  
v_{12}=(-2, 4, -1),
\ea
\eea
which are drawn  in Figure 8. This is exactly the toric data for
the $\Z_3 \times \Z_2$ orbifolded conifold $\CC_{32}$.
The corresponding charge matrix is given by
\bea
Q_{\CM}=\left(
\begin{array}{cccccccccccc}
2& 0& 0& 0& 0& 0& 1& 0& -4& 1& 0& 0\\ -3& 0& 0& 0& 0& 0& -1& 1& 3& 0& 0& 0\\ 
 1& 0& 0& 1& 0& 0& 1& 0& -3& 0& 0& 0\\ -2& 1& 0& 0& 0& 0& -1& 0& 2& 0& 0& 0\\
 -2& 0& 0& 0& 0& 1& 0& 0& 1& 0& 0& 0\\ 1& 0& 0& 0& 0& 0& 0& 0& -2& 0& 1& 0\\
 0& 0& 0& 0& 0& 0& 1& 0& -2& 0& 0& 1\\-1& 0& 0& 0& 1& 0& -1& 0& 1& 0& 0& 0\\
 -1& 0& 1& 0& 0& 0& 1& 0& -1& 0& 0& 0
\end{array}
\right).
\eea
The Fayet-Iliopoulos D-term parameters corresponding to each row is given by
\bea
\left(  
\begin{array}{c}
-3\xi_1 + \xi_3 - 2\xi_4 - 
\xi_5 + 2\xi_6 - 3\xi_7 + \xi_9 - 
   2\xi_{10} - \xi_{11}\\
3\xi_1 - \xi_3 + 2\xi_4 + \xi_5 - 2\xi_6 + 2\xi_7 - \xi_8 - 2\xi_9 + 
\xi_{10}\\
-3\xi_1 - \xi_2 - 2\xi_4 - \xi_5 + \xi_6 - 3\xi_7 - \xi_8 - 
   2\xi_{10} - \xi_{11}\\
2\xi_1 - \xi_3 + \xi_4 - 2\xi_6 + \xi_7 - \xi_8 - 2\xi_9\\
\xi_1 - \xi_2 - \xi_3 - 2\xi_6 - \xi_8 - 2\xi_9 - \xi_{11}\\
-2\xi_1 - \xi_4 - \xi_5 + \xi_6 - 2\xi_7 - \xi_{10} - \xi_{11}\\
-\xi_1 - \xi_4 - \xi_5 - \xi_7 - \xi_{10} - \xi_{11}\\
\xi_1 + \xi_4 + \xi_5 + \xi_7 + \xi_{10}\\
-\xi_1 - \xi_2 - \xi_3 - \xi_4 - \xi_5 - \xi_6 - \xi_7 - \xi_8 - \xi_9 - 
\xi_{10} - \xi_{11}
\end{array}
\right)
\eea

\begin{figure}
\setlength{\unitlength}{0.00063300in}%
\begingroup\makeatletter\ifx\SetFigFont\undefined%
\gdef\SetFigFont#1#2#3#4#5{%
  \reset@font\fontsize{#1}{#2pt}%
  \fontfamily{#3}\fontseries{#4}\fontshape{#5}%
  \selectfont}%
\fi\endgroup%
\begin{picture}(433,2219)(451,-4897)
\thicklines
\put(4801,-2761){\circle*{100}}
\put(4201,-3961){\circle*{100}}
\put(5401,-3361){\circle*{100}}
\put(6601,-2761){\circle*{100}}
\put(3001,-4561){\circle*{100}}
\put(3601,-3361){\circle*{100}}
\put(2401,-3961){\circle*{100}}
\put(1201,-4561){\circle*{100}}
\put(8401,-2761){\circle*{100}}
\put(7201,-3361){\circle*{100}}
\put(6001,-3961){\circle*{100}}
\put(4801,-4561){\circle*{100}}
\put(4801,-2761){\line(-2,-1){3600}}
\put(1201,-4561){\line( 1, 0){3600}}
\put(4801,-4561){\line( 2, 1){3600}}
\put(8401,-2761){\line(-1, 0){3600}}

\put(1051,-4861){$v_1$}
\put(2251,-4261){$v_2$}
\put(3451,-3661){$v_3$}
\put(4651,-3061){$v_4$}
\put(2926,-4861){$v_5$}
\put(4051,-4261){$v_6$}
\put(5251,-3661){$v_7$}
\put(6451,-3061){$v_8$}
\put(4651,-4861){$v_9$}
\put(5926,-4261){$v_{10}$}
\put(7051,-3661){$v_{11}$}
\put(8326,-3061){$v_{12}$}
\end{picture}
\caption{${\bf Z}_3 \times {\bf Z}_2$ orbifold of the conifold $\CC_{23}$}
\end{figure}
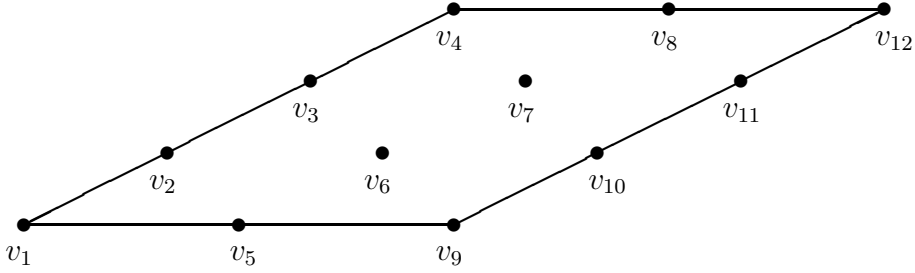

\section{Partial Resolutions}
In order to see  (partial) resolutions of  the singularities in the
formalism used above, we need to turn on the Fayet-Iliopoulos terms.
This will correspond to  triangulations of the convex cone in toric geometry
and moving the center of the moment map in symplectic reduction. 

Before starting the actual discussion, we  make some
observations about the general cases. In \cite{unge} it was considered the
case of generalized conifolds of type $x y = u^{k} v^{k}$ and their
resolutions. Their partial resolutions are conifold singularities, pinch
point singularities and orbifold singularities and are obtained for
different values of the FI parameters. In the T dual picture, D3 branes at
 $x y = u^{k} v^{k}$ singularities transform into $k$ NS branes, $k$ NS$^{'}$
branes on circle together with D4 branes having the circle as one of the
worldvolume coordinates. Partial resolutions of the singularity are
obtained in the T-dual picture by moving one NS brane in the $x^7$
direction (in field theory this means to give expectation values to one
field thus breaking the product of two gauge groups to a diagonal one).
This smoothen the singularity to  $x y = u^{k-1} v^{k}$. By removing a
NS$^{'}$ brane, the singularity is smoothen to  $x y = u^{k} v^{k-1}$.
In \cite{unge}, the starting point was D3 at $x y = u^2 v^2$ singularity
whose T dual contains 2 NS and 2 NS$^{'}$ branes. By removing the two NS branes
one arrives at the conifold singularity, by removing one NS and one NS$^{'}$
one arrives at the conifold and by removing either one NS or one NS$^{'}$ the
pinch point singularity is obtained. This of course means that we resolve
the initial ``worse'' singularity to a ``smoother'' one. By removing NS branes
we have complete control on the spacetime singularity.

In the case of orbifolded conifolds we need to use brane box models
obtained by making two T-dualities. In this case, the resolutions are
obtained either by moving NS and NS$^{'}$ branes with respect to each other or
by opening diamonds at the intersections of the NS and NS$^{'}$ branes. 

The discussion is similar for both types of ${\bf Z}_k \times {\bf Z}_l$
orbifolded conifolds discussed in this paper.
Let us consider the $\Z_2\times \Z_2$ orbifolded conifold case.
From (\ref{QM}), we have a moment map $\mu_{\CM}: \C^9 \to \R^{6}$:
\bea
\mu_{\CM}=\left(
\ba{l}
2|p_3|^2 -2|p_4|^2 -|p_6|^2 +|p_8|^2 -2\xi_1 - \xi_2 - \xi_4 - \xi_5 +\xi_7\\ 
|p_3|^2-|p_4|^2 -|p_6|^2 +|p_7|^2 - \xi_1 + \xi_7\\ 
|p_3|^2-2|p_4|^2 +|p_5|^2 -\xi_1 + \xi_3 + \xi_6 +\xi_7\\
|p_0|^2 -2|p_3|^2 +|p_6|^2+\xi_1 + \xi_2 + \xi_5 + \xi_6\\
|p_2|^2 -2|p_4|^2 +|p_6|^2+\xi_2 + \xi_3 + \xi_6 + \xi_7\\
|p_1|^2 -|p_3|^2-|p_4|^2 +|p_6|^2  +\xi_2 + \xi_6 
\ea
\right)
\eea
where $p_i$ are homogeneous coordinates of $\C^9$.
Then the $\CM$ is the symplectic reduction $\mu_{\CM}^{-1}(0)/U(1)^6$.
>From  the conditions of (\ref{fi-ieq}), Fayet-Iliopoulos parameters
of the resulting $U(1)^6$ gauged linear sigma model satisfy inequalities
\bea 
-2\xi_1 - \xi_2 - \xi_4 - \xi_5 +\xi_7<0,\\ \nonumber
 - \xi_1 + \xi_7 <0,\\ \nonumber
 -\xi_1 + \xi_3 + \xi_6 +\xi_7<0,\\ \nonumber
\xi_1 + \xi_2 + \xi_5 + \xi_6 <0,\\ \nonumber
\xi_2 + \xi_3 + \xi_6 + \xi_7<0.
\eea
But the condition (\ref{fi-ieq}) does not determine the sign of the
last coordinate $\xi_2 + \xi_6$ of the center of the moment map $\mu_{\CM}$.
Notice that the last coordinate of the moment map $\mu_{\CM}$ which is flopped as the sign of 
$\xi_2 +\xi_6$ changes. When $ \xi_2 +\xi_6>0$, it is parameterized
by the homogeneous coordinates $p_3$ and $p_4$. When $ \xi_2 +\xi_6<0$,
 it is parameterized
by the homogeneous coordinates $p_1$ and $p_6$. These two phases are 
topologically
different.
Thus the D-brane vacuum moduli space $\CM$ does have
topologically distinct phases which are  related by a flop transition.
This phenomenon has been observed for orbifold singularities \cite{muto,gre}.
We can see this flop in the toric diagram which is shown in Figure 9 .
\bef
\setlength{\unitlength}{0.00043300in}%
\begingroup\makeatletter\ifx\SetFigFont\undefined%
\gdef\SetFigFont#1#2#3#4#5{%
  \reset@font\fontsize{#1}{#2pt}%
  \fontfamily{#3}\fontseries{#4}\fontshape{#5}%
  \selectfont}%
\fi\endgroup%
\begin{picture}(434,5218)(-2051,-5197)
\thicklines
\put(7202,-4262){\circle*{100}}
\put(8402,-3662){\circle*{100}}
\put(9602,-1862){\circle*{100}}
\put(8402,-2462){\circle*{100}}
\put(7202,-3062){\circle*{100}}
\put(8402,-1262){\circle*{100}}
\put(9602,-662){\circle*{100}}
\put(10802,-62){\circle*{100}}
\put(8401,-1261){\line(-2,-3){1200}}
\put(7201,-3061){\line( 0,-1){1200}}
\put(7201,-4261){\line( 2, 3){1200}}
\put(8401,-2461){\line( 0, 1){1200}}
\put(7127,-4562){$v_2$}
\put(8327,-3962){$v_3$}
\put(9527,-2162){$v_6$}
\put(9527,-962){$v_8$}
\put(10727,-362){$v_9$}
\put(6676,-3211){$v_4$}
\put(8551,-1561){$v_7$}
\put(8551,-2761){$v_5$}
\put(1803,-4262){\circle*{100}}
\put(3003,-3662){\circle*{100}}
\put(4203,-1862){\circle*{100}}
\put(3003,-2462){\circle*{100}}
\put(1803,-3062){\circle*{100}}
\put(3003,-1262){\circle*{100}}
\put(4203,-662){\circle*{100}}
\put(5403,-62){\circle*{100}}
\put(3002,-1261){\line(-2,-3){1200}}
\put(1802,-3061){\line( 0,-1){1200}}
\put(1802,-4261){\line( 2, 3){1200}}
\put(3002,-2461){\line( 0, 1){1200}}
\put(1728,-4562){$v_2$}
\put(2928,-3962){$v_3$}
\put(4128,-2162){$v_6$}
\put(4128,-962){$v_8$}
\put(5328,-362){$v_9$}
\put(1277,-3211){$v_4$}
\put(3152,-1561){$v_7$}
\put(3152,-2761){$v_5$}
\put(6001,-4861){\circle*{100}}
\put(601,-4861){\circle*{100}}
\put(1801,-3061){\line( 2, 1){1200}}
\put(8401,-1261){\line(-2,-5){1200}}
\put(5101,-2461){\line( 1, 0){1500}}
\put(6601,-2461){\makebox(6.6667,10.0000){\SetFigFont{10}{12}{\rmdefault}{\mddefault}{\updefault}.}}
\put(6601,-2461){\makebox(6.6667,10.0000){\SetFigFont{10}{12}{\rmdefault}{\mddefault}{\updefault}.}}
\put(6526,-2461){\vector( 1, 0){ 75}}
\put(6526,-2461){\vector(-1, 0){1350}}
\put(7201,-3061){\makebox(6.6667,10.0000){\SetFigFont{10}{12}{\rmdefault}{\mddefault}{\updefault}.}}
\multiput(1801,-3061)(-7.50000,-11.25000){161}{\makebox(6.6667,10.0000){\SetFigFont{10}{12}{\rmdefault}{\mddefault}{\updefault}.}}
\multiput(601,-4861)(12.06030,6.03015){200}{\makebox(6.6667,10.0000){\SetFigFont{10}{12}{\rmdefault}{\mddefault}{\updefault}.}}
\multiput(3001,-3661)(7.50000,11.25000){321}{\makebox(6.6667,10.0000){\SetFigFont{10}{12}{\rmdefault}{\mddefault}{\updefault}.}}
\multiput(5401,-61)(-12.06030,-6.03015){200}{\makebox(6.6667,10.0000){\SetFigFont{10}{12}{\rmdefault}{\mddefault}{\updefault}.}}
\multiput(3001,-1261)(12.12121,-6.06061){100}{\makebox(6.6667,10.0000){\SetFigFont{10}{12}{\rmdefault}{\mddefault}{\updefault}.}}
\multiput(4201,-661)(0.00000,-13.48315){90}{\makebox(6.6667,10.0000){\SetFigFont{10}{12}{\rmdefault}{\mddefault}{\updefault}.}}
\multiput(4201,-1861)(-12.12121,-6.06061){100}{\makebox(6.6667,10.0000){\SetFigFont{10}{12}{\rmdefault}{\mddefault}{\updefault}.}}
\multiput(3001,-2461)(0.00000,-13.55422){84}{\makebox(6.6667,10.0000){\SetFigFont{10}{12}{\rmdefault}{\mddefault}{\updefault}.}}
\multiput(3001,-3586)(0.00000,-12.50000){7}{\makebox(6.6667,10.0000){\SetFigFont{10}{12}{\rmdefault}{\mddefault}{\updefault}.}}
\multiput(7201,-3061)(-7.50000,-11.25000){161}{\makebox(6.6667,10.0000){\SetFigFont{10}{12}{\rmdefault}{\mddefault}{\updefault}.}}
\multiput(6001,-4861)(12.12121,6.06061){100}{\makebox(6.6667,10.0000){\SetFigFont{10}{12}{\rmdefault}{\mddefault}{\updefault}.}}
\multiput(7201,-4261)(12.12121,6.06061){100}{\makebox(6.6667,10.0000){\SetFigFont{10}{12}{\rmdefault}{\mddefault}{\updefault}.}}
\multiput(8401,-3661)(7.50000,11.25000){321}{\makebox(6.6667,10.0000){\SetFigFont{10}{12}{\rmdefault}{\mddefault}{\updefault}.}}
\multiput(10801,-61)(-12.06030,-6.03015){200}{\makebox(6.6667,10.0000){\SetFigFont{10}{12}{\rmdefault}{\mddefault}{\updefault}.}}
\multiput(8401,-1261)(12.12121,-6.06061){100}{\makebox(6.6667,10.0000){\SetFigFont{10}{12}{\rmdefault}{\mddefault}{\updefault}.}}
\multiput(9601,-1861)(0.00000,13.48315){90}{\makebox(6.6667,10.0000){\SetFigFont{10}{12}{\rmdefault}{\mddefault}{\updefault}.}}
\multiput(8401,-2461)(12.12121,6.06061){100}{\makebox(6.6667,10.0000){\SetFigFont{10}{12}{\rmdefault}{\mddefault}{\updefault}.}}
\multiput(8401,-2536)(0.00000,-13.55422){84}{\makebox(6.6667,10.0000){\SetFigFont{10}{12}{\rmdefault}{\mddefault}{\updefault}.}}
\put(451,-5161){$v_1$}
\put(5851,-5161){$v_1$}
\put(5626,-2311){\mbox{Flop}}
\end{picture}
\caption{A flop transition between different phases}
\enf

For special values of $\xi_i$, 
there are several singularity types. Of course, we get the orbifolded conifold
$\CC_{22}$ when all $\xi$ are zero. But the 
singularity becomes partially resolved,
when fields get expectation values in terms of the FI parameters.
One of the most interesting case is when we give 
expectation values to the fields $A_{2i}, i= 1 ,\cdots, 4$. 
This region corresponds to $\xi_5 + \xi_1 =\xi_6 + \xi_2=\xi_7 + \xi_3
=\xi_8 + \xi_4$. Hence the last three coordinates of the 
center of the moment map $\mu_{\CM}$ are zeros. Thus one can see 
that the lower left half triangle of the toric diagram  will not be
triangulated. So we will have an orbifold singularity  $\C^3/\Z_2 \times
\Z_3$ for generic values of $\xi_i$ under these circumstances (Figure 10).
\bef
\setlength{\unitlength}{0.00053300in}%
\begingroup\makeatletter\ifx\SetFigFont\undefined%
\gdef\SetFigFont#1#2#3#4#5{%
  \reset@font\fontsize{#1}{#2pt}%
  \fontfamily{#3}\fontseries{#4}\fontshape{#5}%
  \selectfont}%
\fi\endgroup%
\begin{picture}(33,5210)(1,-5488)
\thicklines
\put(3601,-5161){\circle*{100}}
\put(4801,-4561){\circle*{100}}
\put(6001,-3961){\circle*{100}}
\put(7201,-2161){\circle*{100}}
\put(6001,-2761){\circle*{100}}
\put(4801,-3361){\circle*{100}}
\put(6001,-1561){\circle*{100}}
\put(7201,-961){\circle*{100}}
\put(8401,-361){\circle*{100}}
\put(6001,-1561){\line(-2,-3){2400}}
\put(3601,-5161){\line( 2, 1){2400}}
\put(6001,-3961){\line( 0, 1){2400}}
\multiput(6076,-3886)(4.99426,7.49139){470}{\makebox(6.6667,10.0000){\SetFigFont{10}{12}{\rmdefault}{\mddefault}{\updefault}.}}
\multiput(6001,-1561)(8.05369,4.02685){299}{\makebox(6.6667,10.0000){\SetFigFont{10}{12}{\rmdefault}{\mddefault}{\updefault}.}}
\multiput(7126,-961)(9.37500,0.00000){9}{\makebox(6.6667,10.0000){\SetFigFont{10}{12}{\rmdefault}{\mddefault}{\updefault}.}}
\put(7201,-961){\makebox(6.6667,10.0000){\SetFigFont{10}{12}{\rmdefault}{\mddefault}{\updefault}.}}
\multiput(7201,-961)(0.00000,-9.02256){134}{\makebox(6.6667,10.0000){\SetFigFont{10}{12}{\rmdefault}{\mddefault}{\updefault}.}}
\multiput(7201,-2161)(-8.05369,-4.02685){150}{\makebox(6.6667,10.0000){\SetFigFont{10}{12}{\rmdefault}{\mddefault}{\updefault}.}}
\multiput(6001,-1561)(8.05369,-4.02685){150}{\makebox(6.6667,10.0000){\SetFigFont{10}{12}{\rmdefault}{\mddefault}{\updefault}.}}
\put(4726,-4861){$v_2$}
\put(3451,-5461){$v_1$}
\put(5926,-4261){$v_3$}
\put(4726,-3661){$v_4$}
\put(5926,-3061){$v_5$}
\put(7126,-2461){$v_6$}
\put(5926,-1861){$v_7$}
\put(7126,-1261){$v_8$}
\put(8326,-661){$v_9$}
\end{picture}
\caption{A phase with orbifold singularity}
\enf
The configuration of D3 branes at this singularity is T-dual to a $2 \times 2$
brane box with trivial identification of the unit cell. In the language of 
\cite{karch}, giving expectation values to the fields $A_{2i}$, i.e.
going to a baryonic branch, means to 
rotate the diamonds which lie at the intersections of the NS and NS'
branes. 
One can have  similar  discussions  for the ${\bf Z_2} \times {\bf Z_3}$
orbifolded conifold. 

\section{Conclusions} 

In this paper,  we have used the toric geometry and Witten's
gauged linear sigma model  to 
identify the Higgs moduli space of the field theory  on the world volume of
branes
at  the orbifolded conifold singularity of type $\CC_{kl}$ which is
a $\Z_k \times \Z_l$ quotient of the conifold $xy -uv = 0$. 
We have shown that the Higgs moduli space does have phases related by 
a flop transition and topology change can occur. 
It is also observed that the orbifold singularity can
 be obtained as one of the phases of 
the Higgs moduli space. In field theory, this corresponds to  
 giving expectation value to some
hypermultiplets. 

Moreover, we have studied a  correspondence between brane
configurations and brane at singularities for the case of orbifolded
conifolds of type $\CC_{kl}$.

\section{Acknowledgments}
We would like to thank Dieter Lust and Andreas Karch for very important
discussions. The work of K. Oh is supported in part
by NSF grant PHY-9970664. R. Tatar would like to thank the Department of Mathematics
at U. Missouri-St. Louis for hospitality.

\end{document}